# On the Unified View of Extragalactic Sources based on their Broadband Emission Properties

E.U. Iyida[*1.2], I.O Eya[1.3], F. C. Odo[1.2]

1 Astronomy and Astrophysics Research Group, Department of Physics/Astronomy, University of Nigeria, Nsukka,
2 Department of Physics and Astronomy, Faculty of Physical sciences, University of Nigeria, Nsukka
3 Department of Science Laboratory Technology, Faculty of Physical sciences, University of Nigeria, Nsukka,
*email: uzochukwu.iyida@unn.edu.ng

**Abstract**

Our understanding of the unification of jetted AGNs has advanced greatly as the size of extragalactic sources increased. In the present paper, based on the large sample of radio sources, we compiled 680 blazars (279 FSRQs and 401 BL Lacs) from the 3FGL sample and 64 Seyfert galaxies (3 Narrow-line, 34 and 27 regular Seyfert 1 and Seyfert 2 respectively) from the INTEGRAL/IBIS survey to statistically test the relationship between Seyfert galaxies and the blazar samples of FSRQs and BL Lacs. We compute the synchrotron (*SS*), Compton (*CS*) and inverse Compton (*IC*) continuous spectra from the low energy components of radio to X-ray, radio to γ-ray and the high energy component of X-ray to γ-ray bands, respectively. Results show from the distributions of the continuous spectra that Seyfert galaxies form the tail of the distributions, suggestive of similar underlying history and evolution. A two-sample Kolmogorov-Smirnov test (*K-S* test) of the continuous spectra showed that Seyfert galaxies differ from BL Lacs and FSRQs in the low energy components of the spectra, while there is no clear difference between them in the high energy component, which implies that high energy emissions in Seyfert galaxies, BL Lacs and FSRQs may be as a result of the same emission mechanism. There is a regular sequence of the distributions on *SS* – *CS* and *IC* –*CS* planes in each individual subsample. Linear regression analyses of our sample yield significant positive correlations ($r \geq 0.60$) between *SS* – *CS* and *IC* – *CS* data. This upturns into an anti-correlation ($r > -0.60$) in *IC* – *SS* data. These results are not only consistent with unified scheme for blazars but also show that Seyfert galaxies can be unified with the classical radio-loud AGNs counterparts.

# 1 Introduction

Active galactic nuclei (AGNs) are dense cores of extragalactic objects characterized by high luminosity coming from non-stellar sources. The morphology of AGNs is indeed complex with emissions at different frequencies originating from various spatial locations or components (Ho, 2008; Heckman & Best, 2014). This includes, the continuum radiation in the radio up to optical/ultraviolet (UV), X-rays and γ-rays coming from the relativistic jet, the cold accretion disc and the hot accretion flow in the case of low-luminosity sources, respectively. AGNs have shown to exhibit bimodality in the distribution of their radio-loudness parameter ($R_L$, defined as the ratio of the flux density at 5 GHz to the flux at *B*-band), classifying these sources into radio-quiet ($R_L \leq 10$) and radio-loud ($R_L > 10$) (Kellermann et al. 1989; Xu et al., 1999; Zhang et al., 2021). The physical origin of their bimodal nature is mostly ascribed to the differences in the fundamental parameters of the galactic system such as the accretion rate, emission mechanisms and the spin rate of the black hole (McLure & Jarvis, 2004). Thus, the degree of the radio-loudness can be used to distinguish between the properties of these extragalactic sources.

In addition, AGNs are known to be either jetted or non-jetted considering the intensity of the powerful relativistic jet which is weak in radio-quiet and very strong in radio-loud AGNs. Blazars are exceptional subsets of jetted, radio-loud AGNs whose ultra-relativistic jets are aligned close to an observer's line of sight. (see, Blandford & Rees, 1978; Urry & Padovani, 1995). They have unique observational properties such as rapid and high amplitude variability, optical polarization, large and variable luminosity, apparent superluminal motion and very active GeV to TeV γ-ray emissions (Fan et al. 2011; Yang et al., 2018; Abdollahi et al., 2020; Ajello et al., 2020). Blazars are traditionally classified into two sub-classes: the BL Lacertae Objects (BL Lacs) and flat spectrum radio quasars (FSRQs). The main distinction between the two subgroups is the intensity of their optical emissions. Whereas the emission lines of FSRQs are very strong, that of the BL Lacs are faint and in most cases absent. Also, the luminosity of the broad-line region (BLR) is distinct between the blazar subsets (Ghisellini & Celotti, 2001). The ratio of broad emission line luminosity to Eddington luminosity ($L_{BLR}/L_{Edd}$) has been used to classify these sources (see, e.g., Ghisellini et al., 2011). FSRQs have $L_{BLR}/L_{Edd} \geq 5 \times 10^{-4}$ while BL Lacs have $L_{BLR}/L_{Edd} < 5 \times 10^{-4}$. This classification may imply that FSRQs and BL Lacs have different emission properties and accretion regime (Sbarrato et al. 2014). Nevertheless, BL Lacs are further classified based on the value of their synchrotron peak frequencies ($v_{s,peak}$). Thus, there are Low Synchrotron Peaked (LSPs), Intermediate Synchrotron Peaked (ISPs) and High Synchrotron Peaked (HSPs) BL Lacs with peak frequencies of $\log \nu_{s,peak} < 14 \,(\text{Hz})$, $14 \leq \log \nu_{s,peak} \leq 15 \,(\text{Hz})$ and $\log \nu_{s,peak} > 15 \,(\text{Hz})$ respectively (see, Ackermann et al., 2015; Fan et al. 2016; Ajello, 2020).

A great effort of research on extragalactic sources has been directed to the development of a unified scheme in which the observed properties of AGNs subclasses could be explained as similar objects seen at different orientation angles to the line of sight. In this context of the unified scheme, all extragalactic radio sources are assumed to possess the same basic morphological features and exhibit similar physical phenomena, the observations of which at different orientation angles give rise to the different classes (Antonucci, 1993). Accordingly, in the perspective of broadband spectral energy distributions (SEDs), the unification scheme of blazars is a kind of arrangement such that the frequency at the peak of synchrotron emission decreases with increasing synchrotron luminosity (Sambruna et al., 1996; Fossati et al. 1997; 1998; Ghisellini & Tavecchio, 2008). This is called the unification model via blazar sequence. This type of connection among the blazar subclasses has been topical in blazar astronomy for over two decades and a number of results has been recorded (e.g. Cosmastri et al., 1997; Fossati et al., 1998; Cellone et al., 2007; Ackermann et al. 2011; Gaur et al., 2010; Urry, 2011; Fan et al. 2016; Iyida et al. 2019; Pei et al. 2019; 2020a). Nevertheless, the phenomenon of blazar astronomy is sometimes explained in the context of relativistic beaming effect (see, Kollgaard, 1994; Wu et al. 2014) due to the fact that the radiations from the jets are assumed to always be directed close to an observer's line of sight. This implies that there is a link among the various types of blazar. Earlier investigations appear to offer meaningful indication that FSRQ samples can metamorphose into BL Lacs via luminosity evolution (Sambruna et al., 1996; Odo et al., 2014; Iyida et al. 2021a). In particular, an evolutionary scenario that relates the FSRQs and BL Lacs in terms of the decrease in black hole accretion power with time has been proposed using the analysis from the self-synchrotron and Compton scattered external radiation from blazar jets (see, Chiang & Bottcher, 2002; Zhang et al., 2012; Odo et al., 2017). These scenarios propose trends in the division of the properties between the FSRQs and BL Lac subclasses from the lower part of the continuum to the extreme (e.g. Ghisellini et al., 1998; Giommi et al., 2012). Similarly, the SEDs of these objects show prominent stability which agrees with the unification scheme confirming the validity of blazar sequence, though the mechanism powering the link among the blazar subclasses is still unclear as their individual emission continuum forms are significantly different (Sambruna et al., 1996; Ghisellini et al., 1998). Even though, the pattern of synchrotron as well as inverse Compton scattering by the leptonic model in ultra-relativistic jets that points close to an observer's line of sight has been used in modeling the SEDs of blazars, many fundamental issues remain, including the properties of these sources in the low and high energy components. Conversely, there seems to be a general agreement that appears to support a unification scheme with interesting results, indicating orientation links between blazar subclasses and radio galaxies (Fan et al. 2011; Pei et al. 2019, Iyida et al. 2020). Precisely, latest studies using the *Fermi*-LAT sample suggest that blazar unification can be extended to radio galaxies by simply invoking an orientation sequence (see, e.g. Pei et al., 2020a; Iyida et al., 2021a).

Meanwhile, since the launch of *Fermi* Large Area Telescope (*Fermi*-LAT), numerous new high-energy γ-ray sources have been detected, thereby, increasing our understanding of γ-ray blazars and offering new prospects of studying the γ-ray production mechanisms (see, Ajello et al., 2020; Abdolahhi et al. 2020). A general indication is that γ-ray emissions of powerful blazars are mainly produced within the BLR from external inverse Compton interaction, EC (see, Ghisellini & Madau, 1996; Dermer et al. 2009; Paliya et al. 2015) whereas for the high synchrotron peaking blazars, it results due to the synchrotron self-Compton mechanism, SSC, (e.g., Maraschi et al. 1992; Finke et al. 2008). Thus, recent analyses of γ-ray loud blazar samples suggest a close relationship between multi-wavelength properties of blazars and their unification scheme (Savolainen et al. 2010; Iyida et al. 2021b). Similarly, investigations of γ-ray data appear to provide evidence that orientation unification scheme is essential in explaining the variation of γ-ray emissions from blazars (e.g. Chen et al. 2016; Odo and Aroh, 2020). This is supported by many correlations between the γ-ray flux and multi-wavelength properties of γ-ray loud sample of extragalactic sources (see, Giommi et al. 2013). In particular, the condition that γ-ray sources have small linear sizes as deduced from γ-ray variability suggest that γ-ray emissions originate from the jets, in the same manner as the low energy emissions. Also, the spectral flux of these sources has been found to depend strongly on the viewing angle, thus, suggesting a systematic trend in the variation of γ-ray flux of the extragalactic sources from low to high γ-ray flux, which can represent a form of unification scheme for jetted AGNs.

In another development, a subset of jetted AGNs that is understood to be imperative in the unification scheme is the radio-quiet Seyfert galaxy (see, e.g, Foschini et al. 2015; Pei et al. 2020a). More evidence has shown that these galaxies which host the powerful ultra-relativistic jets with small black hole mass could develop to FSRQs or the conventional FR IIs (see, Abdo et al., 2009; Paliya et al., 2019; Sun et al. 2015; Foschini et al. 2015; 2017). In general, it is argued that Seyfert galaxies display some properties (e.g. large amplitude, compact radio cores, very high brightness temperatures, spectral variability, increased continuum emission, flat X-ray and γ-ray spectra e.t.c.) that are blazar-like (see, Yuan et al., 2008; Abdo et al. 2009; Foschini et al., 2015; 2017), indicating that they can be unified with blazars. Furthermore, the similarities between the nuclei of Seyfert galaxies and the radio-loud AGNs have also been discussed earlier (see, e.g. Dahari & De Robertis, 1988; Ho & Ulvestad, 2001; Falcke et al. 2000). Thus, numerous efforts have been recorded to demonstrate a continuity in overall distributions of the observed properties of the Seyfert galaxies and the traditional radio-loud AGNs. Although a vast majority of these regular galaxies are known to be traditionally radio-quiet objects, the radio-loud galaxies are believed to harbor powerful relativistic jets with extended radio structures comparable to Seyfert galaxies (Liu & Zhang, 2002; Doi et al. 2012; Mathur *et al*, 2012). Thus, the new unified scheme of AGNs hypothetically, embraces these blazar-like galaxies as young jetted counterparts of traditional radio-loud AGNs or instead a part of a larger AGN subclass observed under a specific geometry and inclination of the line of sight (Singh & Chand, 2018). Specifically, Pei et al., (2019; 2020a) alluded to a possible unification of FSRQs with Seyfert galaxies via relativistic beaming and source orientation as the authors found that the core-dominance parameter (defined as the ratio of core-to-lobe flux) is strongly anti-correlated with luminosity in a sample of FSRQs and Seyfert galaxies.

Therefore, the challenge of searching for jetted AGNs in the radio-quiet class and considering linking them with the radio-loud counterparts in the low and high energy components of SEDs could be a worthwhile exploration and is somewhat the motivation for the current investigation. In this paper, we perform a statistical study of the properties of blazars and these Seyfert galaxies in order to gain insight into the nature of their broadband emissions. Our particular aim is to investigate the broadband emission mechanisms of Seyfert galaxies and blazar subsets in the light of unification scheme. The paper is presented in this way: section 2 discusses the methodology and basic assumptions while section 3 describes the sample selection and analyses. Discussion and conclusion of this paper are presented in Sections 4 and 5, respectively.

## 2 Methodology and Basic Assumptions

The low energy component of SEDs is very well known to result from the synchrotron emission coming from ultra-relativistic electron in the jet that is directed at small angle to the line of sight. Nonetheless, in the high energy end, there are two models: the simple leptonic and hadronic models. Within the background of the simple leptonic model, high energy emissions result from the Compton scattering of low frequency photon away from the relativistic electrons that peak in radio to UV/X-ray bands, while the inverse Compton scattering by similar number of electron produces the highest energy component that peak at X-ray/ γ-ray energies (Sikora & Madejski, 2001; Krawczynski et al. 2004; Bottcher et al., 2007). Thus, the continuous emission of blazars is observed to vary over many orders of magnitude in spectrum. Some authors modeled some parameters that depend on the broadband properties to study blazar unification scheme (see, Nalewajko and Gupta, 2017; Iyida et al., 2020). Of these proxy parameters often used is the composite spectral indices ($\alpha_{ij}$), defined mainly between two wavebands (see, e.g. Iyida et al., 2020) as:

$$\text{Composite Spectral indices}\,(\alpha_{ij}) = -\frac{\log\left(\frac{L_i}{L_j}.F\right)}{\log\left(\frac{vi}{vj}\right)} \qquad (1)$$

where *i* and *j* signify two arbitrary wavebands, *L* is the monochromatic luminosity of the objects while *v* and *F* are the observed frequency and the total *k*-correction factor respectively. Even though there is substantial similarity of synchrotron, Compton and inverse Compton emissions from radio to γ-ray bands for large sample of blazars, for convenience, we refer the radio to X-ray continuous spectrum as synchrotron spectrum (*SS*) because it arises from the synchrotron radiation (e.g. Finke, 2013), radio to γ-ray spectrum as Compton spectrum (*CS*) since it is produced through Compton up-scattering of soft seed photons by the same population of relativistic electrons that also produces the synchrotron emission at lower frequencies (synchrotron self-Compton) (see, e.g Sikora & Madejski, 2001; Bottcher et al., 2007) and the X-ray to γ-ray spectrum as inverse Compton spectrum (*IC*) because it peaks at GeV – TeV energies (Bottcher et al., 2007; Krawczynski et al. 2004; Finke, 2013). Though, it is generally understood that high synchrotron peaking BL subclasses with up to GeV energies are evolving from recently

4observed data (see e.g. Foffano et al., 2019), suggesting that the GeV emissions are not purely of Compton spectra, the statistical information of these sources are unavailable. In fact, very few objects (55) have been reported to be detected in GeV to TeV energy in the third *Fermi*-LAT (3FGL) blazar catalogue (Acero et al. 2015; Ackermann et al. 2015). However, the high energy (GeV up to TeV) Compton spectra of these objects have not been explicitly modeled (see, e.g. Ackermann et al. 2015). Therefore, significant number of these γ-ray-loud sources rely largely on *Fermi*-LAT data, thus, the assumed modeling is valid.

The photon number per unit energy (E) of extragalactic source with spectral index ($\Gamma$) is expressed as

$$N_0 E^{-\Gamma} = \frac{dN}{dE} \tag{2}$$

where the $N_0$ is initial flux obtained by the integration of equation (2). This gives

$$N_0 = N \frac{E_L E_U}{E_L - E_U} \tag{3}$$

with $E_L - E_U$ being the low and high energy range given differently for the wavebands. The value of $N$ is the integral photon flux in units of photons cm$^{-2}$/s in the energy range of $E_L - E_U$. For instance, the broadband luminosity at 1 GeV can be expressed as

$$L = 1.91 \times 10^{47} d_L^2 \frac{N(1-\Gamma)}{E_U^{1-\Gamma} - E_L^{1-\Gamma}} \cdot E^{2-\Gamma} \tag{4}$$

where $d_L$ is the luminosity distance given as $d_L = H_o^{-1} \int \left[ (1+z)^2 (1+\Omega_m z) - z(2+z)\Omega_\Lambda \right]^{-1/2} dz$. This broadband luminosity can be calculated from the expression

$$L = 4\pi d_L^2 S_\nu^{obs} (1+z)^{\Gamma-2} \tag{5}$$

with $S_\nu^{obs}$ being the observed total flux ranging from low (100 MeV) to the high (100 GeV) energy end. Thus, the broadband luminosity can be obtained if the photon spectral index is in the energy range of 100 MeV to 100 GeV using,

$$L = 1.91 \times 10^{47} d_L^2 \frac{E_L E_U}{E_L - E_U} \cdot E^{2-\Gamma} \tag{6}$$

Consequently, if there are available information on different wavebands, the relationship between blazar subsets and Seyfert galaxies using the correlations among their spectra at different energy ranges can be investigated.

## 3 Data Selection

The data sample for this paper is from the third catalogue of AGNs (3FGL) detected by *Fermi* Large Area Telescope (*Fermi*-LAT) and compiled by Acero et al., (2015) and Ackermann et al., (2015). There is a considerable similarity in the data samples presented by the two groups of authors, and thus, for uniformity, we obtained the main information, including the SED classifications, from Acero et al. (2015). First, we take a sample of 1081 blazars with defined optical identifications comprising 620 BL Lacs and 461 FSRQs and collected their radio (*Lr*), X-ray (*Lx*) and γ-ray (*Lγ*) monochromatic luminosities from the available $1.40 \times 10^9$ Hz, $2.52 \times 10^{17}$ Hz and $2.52 \times 10^{23}$ Hz data respectively. We did not consider the unclassified blazars (BCU) from either sample so as to have a clear sample. The clean sample contains 680 blazars (279 FSRQs and 401 BL Lacs) with comprehensive information on the three monochromatic luminosities under consideration.

However, to enable proper investigation of the unification of the extragalactic sources through their broadband emission properties, we considered and included 64 jetted Seyfert galaxies (3 Narrow-line, 34 and 27 regular Seyfert 1 and Seyfert 2 respectively) from the INTEGRAL/IBIS survey with measured radio and X-ray monochromatic luminosities from Chang et al. (2021) at the same observing frequencies of $1.40 \times 10^9$ Hz and $2.52 \times 10^{17}$ Hz respectively while the γ-ray luminosity was calculated from equation 5 using the information published in Pei et al. (2020b). For each of the objects in the whole sample of 744 sources (680 blazars and 64 Seyfert galaxies), we calculated: s*ynchrotron*, *inverse Compton and Compton* continuous spectra from the available monochromatic properties. Throughout the paper, the standard cold dark matter (Λ-CDM) cosmology was adopted with Hubble's constant $H_0$ = 72 kms⁻Mpc$^{-1}$ and $\Omega_0 = \Omega_m + \Omega_\Lambda$, ($\Omega_m$ = 0.30; $\Omega_\Lambda$ = 0.70). All relevant data are adjusted based on this concordance cosmology. In the case of statistical analyses, the Pearson Product Moment correlation coefficient (*r*) is used to determine the degree of relationship between these emission parameters using python and MatLab programming softwares.

**4 Analysis and Results**

In this section, we discuss the statistical comparison of the distributions and correlations of synchrotron, inverse Compton and Compton continuous spectra of Seyfert galaxies, FSRQs and BL Lacs in the framework of the unification scheme.

**4.1 Distributions of the continuous spectra of our sample**

The comparison of the histogram distributions of the parameters of our sample is not only necessary in studying their emission mechanisms but also essential in understanding the relationship among the various subclasses of these sources. However, for the purpose of the unification of extragalactic sources, these parameters are considered individually in our sample. The distribution of synchrotron spectrum (*SS*) and the cumulative probability of our sample are shown respectively in Figures 1a and 1b. Obviously, the distribution of these objects is continuous, with the Seyfert galaxies located at the lowest part of the spectrum; FSRQs appear to have larger values of synchrotron spectrum in the data as shown in Figure 1a. It can also be observed from the figure that FSRQs and Seyfert galaxies are not properly fitted to normal distribution as they skew to the left and right with values of -0.07 and 0.03 respectively. However, HSPs have unimodal configuration peaking at -3.56 while ISPs and LSPs have rather flat distributions. Nevertheless, the distributions give mean values of ~ -4.20 ± 0.05 for Seyfert galaxies, -3.04 ± 0.08 for HSPs, -3.39 ± 0.04 for ISPs, -3.62 ± 0.08 for LSPs and -2.40 ± 0.03 for FSRQs. Nonetheless, a two-sample Kolmogorov-Smirnov (*K-S*) test between the distributions of *SS* for Seyfert galaxies, FSRQs and individual subclasses of BL Lacs indicate that though, the observed average sequence exists (indicative of a unified scheme for these sources) still, the distributions in individual subsamples are significantly different, indicating that there are many diverse intrinsic properties among the subsamples in the low energy component. The *K-S* test result is shown in Table 1 where *n* is the number of subsample, $d_{max}$ is the separation distance while *p* is the chance probability.

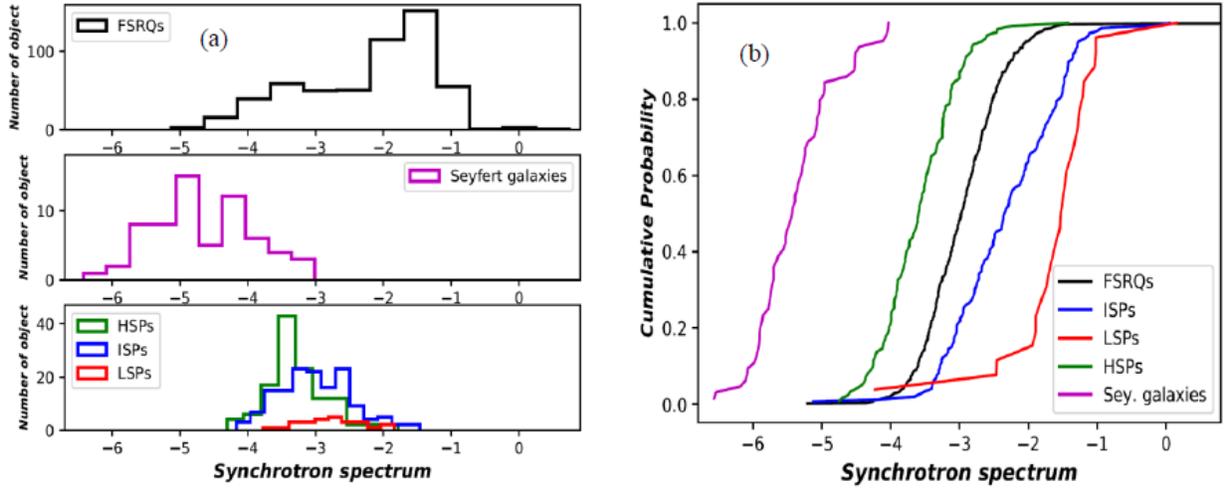

**Figure 1:** Histogram showing comparison of (a) synchrotron spectrum of FSRQs, Seyfert galaxies and BL Lac subclasses (b) cumulative distribution function of synchrotron spectrum of the sample.

**Table 1:** Results of *K-S* test of *SS* of our sample

| **Parameter** | *Samples* | *n* | *d$_{max}$* | *p* |
|---|---|---|---|---|
| Synchrotron spectrum | Seyfert galaxies - HSP | 64 -138 | 0.39 | $5.20 \times 10^{-10}$ |
| Synchrotron spectrum | Seyfert galaxies - ISP | 64 -133 | 0.66 | $1.12 \times 10^{-12}$ |
| Synchrotron spectrum | Seyfert galaxies - LSP | 64 -130 | 0.71 | $2.32 \times 10^{-15}$ |
| Synchrotron spectrum | Seyfert galaxies - FSRQs | 64 - 279 | 0.85 | $4.25 \times 10^{-23}$ |

To further support the unification of Seyfert galaxies and blazar subclasses, we show in Figure 2, the histogram comparing the distributions of inverse Compton spectrum (*IC*) for FSRQs, Seyfert galaxies and BL Lac subclasses. While Seyfert galaxies and blazar subclasses occupy similar spaces and generally overlap in inverse Compton spectrum with up to 3 orders of magnitude implying historic connection, BL Lacs and FSRQs span the entire range with mean values of 1.05 ± 0.06, -1.07 ± 0.03, -1.02 ± 0.02 and 0.14 ± 0.04 for FSRQs, LSPs, ISPs and HSPs, respectively and 0.62 ± 0.03 for Seyfert galaxies. However, we noted that for all the distributions, there is continuity with no distinct dichotomy between the subclasses in such a way that is in agreement with the unification scheme, signifying that the populations of the sample are inherently related. A *K-S* test was done on the inverse Compton spectrum data and the results are presented in Table 2. We observed from the cumulative distribution function in Figure 2b that the overlap is very obvious mainly for LSPs and Seyfert galaxies, signifying that FSRQs, HSPs and ISPs are statistically different while LSPs and Seyfert galaxies share a similar emission mechanism in inverse Compton spectrum. In view of this *K-S* test, Seyfert galaxies can be unified with the blazar subclasses.

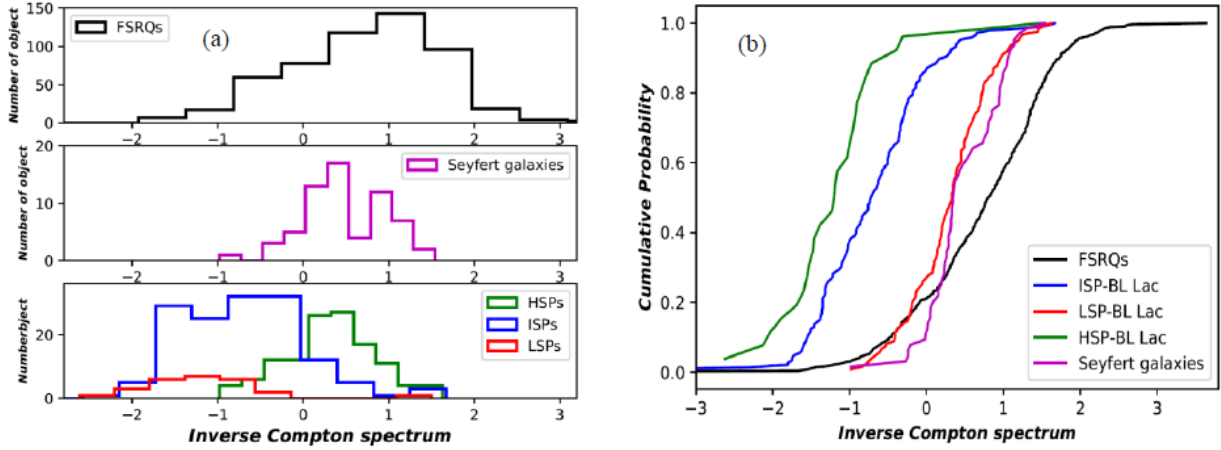

**Figure 2:** Histogram showing comparison of (a) inverse Compton spectrum of FSRQs, Seyfert galaxies and BL Lac subclasses (b) cumulative distribution function of the sample

**Table 2:** Results of *K-S* test of *IC* of our sample

| **Parameter** | *Samples* | *n* | *$d_{max}$* | *p* |
|---|---|---|---|---|
| Inverse Compton spectrum | Seyfert galaxies - HSP | 64 -138 | 0.62 | $2.02\times10^{-17}$ |
| Inverse Compton spectrum | Seyfert galaxies - ISP | 64 -133 | 0.56 | $3.35\times10^{-13}$ |
| Inverse Compton spectrum | Seyfert galaxies - LSP | 64 -130 | 0.48 | $6.12\times10^{-16}$ |
| Inverse Compton spectrum | Seyfert galaxies -FSRQs | 64- 279 | 0.53 | $4.43\times10^{-14}$ |

The distributions of the objects in Compton spectrum (*CS*) is shown in Fig. 3. Evidently, the distribution of the subsamples is continuous, with FSRQs occupying the highest Compton spectrum while Seyfert galaxies form the tail of the spectrum, suggesting that BL Lacs and Seyfert galaxies are purely low energy component of AGNs. Meanwhile, the mean values of our subsamples from the distributions give ~ -2.84 ± 0.06 for FSRQs, -2.84 ± 0.08 for LSPs, -2.41 ± 0.05 for ISPs, -2.22 ± 0.04 for HSPs and -4.34 ± 0.02 for Seyfert galaxies. Further statistical investigations indicate that FSRQs do not have any normal distribution as it shifts to the left with skewness (*μ*) in the range - 0.03 ≤ μ ≤ 0.05. However, Seyfert galaxies and BL Lacs have near normal distribution with *μ* in the range - 0.01 ≤ μ ≤ 0.02. A *K-S* test was carried out on the *CS* data. Results show that generally, there is approximately zero probability that there is a fundamental difference between the underlying distributions of these objects in Compton spectrum. The cumulative probability function is shown in Fig. 3b while the *K-S* test result is shown in Table 3. From the distributions and the *K-S* test result, we can find that the mean values follow the sequence $\langle CS \rangle|_{\text{Seyfert galaxies}} < \langle CS \rangle|_{\text{HSPs}} < \langle CS \rangle|_{\text{ISPs}} < \langle CS \rangle|_{\text{LSPs}} < \langle CS \rangle|_{\text{FSRQs}}$ indicative of Seyfert galaxy - BL Lacs – FSRQs unification. This result can be interpreted to mean that in terms of Compton spectrum, different subclasses of BL Lacs may exist and that continuous spectra can be used to investigate the unification scheme.

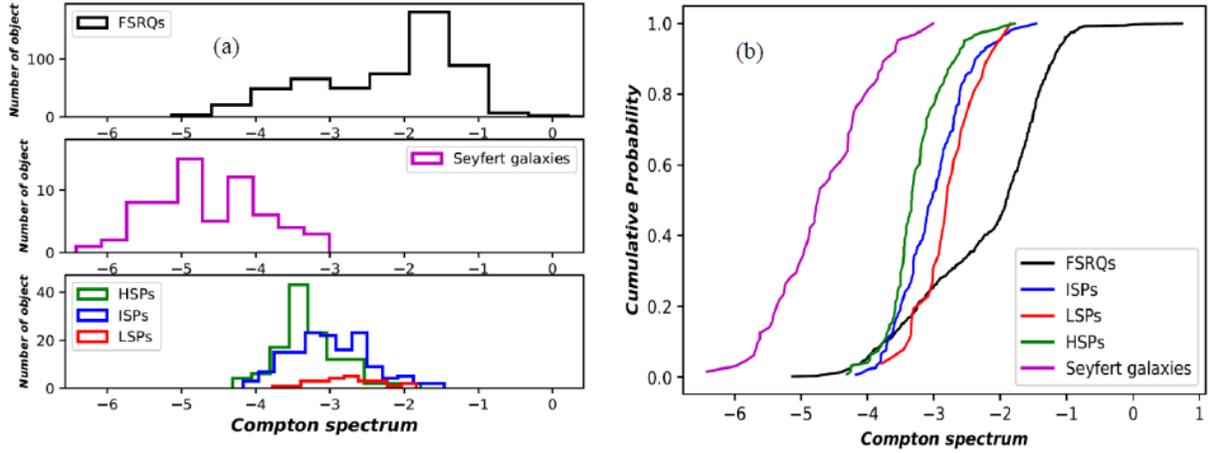

**Figure 3: Histogram showing comparison of (a) Compton spectrum of FSRQs, Seyfert galaxies and BL Lac subclasses (b) cumulative distribution function of the sample**

**Table 3:** Results of *K-S* test of *CS* of our sample

| Parameter | *Samples* | *n* | *d$_{max}$* | *p* |
|---|---|---|---|---|
| Compton spectrum | Seyfert galaxies - HSP | 64 -138 | 0.35 | $3.09 \times 10^{-18}$ |
| Compton spectrum | Seyfert galaxies - LSP | 64 -133 | 0.62 | $2.32 \times 10^{-21}$ |
| Compton spectrum | Seyfert galaxies - LSP | 64 -130 | 0.70 | $7.96 \times 10^{-16}$ |
| Compton spectrum | Seyfert galaxies - FSRQs | 64 - 279 | 0.84 | $9.02 \times 10^{-14}$ |

**4.2 Correlations among the continuous spectra and the unification scheme**

The unified scheme of AGNs predicts the existence of correlations between emissions in different wavebands. The study of these correlations can provide valuable information about the emission mechanisms and the connection between the extragalactic sources. We analyzed whether the correlations among the continuous spectra for Seyfert galaxies, FSRQs and BL Lacs are in agreement with the prediction of the unified scheme. The scatter plot of *SS* as a function of *CS* is shown in Figure 4. The Seyfert galaxies and FSRQs align in a sense that is consistent with the predicted unification scheme. Meanwhile, two distinct groups of objects can be identified, namely, Seyfert galaxies - FSRQs and the BL Lacs. There is a clear positive correlation of *SS* – *CS* data for each of the groups with the Seyfert galaxies - BL Lacs - FSRQs sequence obvious in the plot. This indicates that on the *SS* – *CS* plane, these objects have trends suggestive of a unified scheme. Table 4 shows the results of linear regression analysis of the samples. The slope *k*, intersection $k_0$, correlation coefficient *r* and chance probability *p* and their errors are all listed in the table.

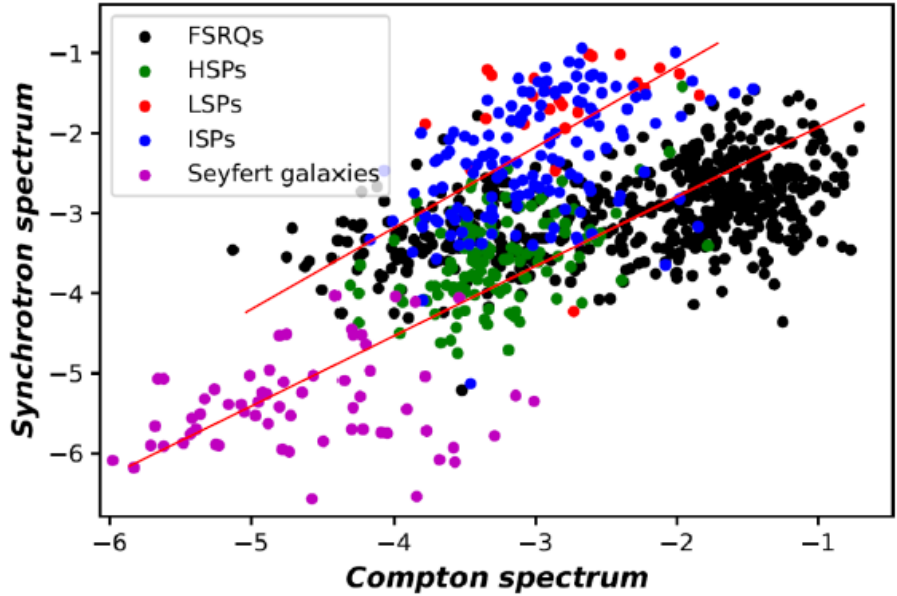

**Figure 4: Plot of synchrotron spectrum against Compton spectrum for FSRQs, Seyfert galaxies and BL Lacs**

Table 4: Results of linear regression fitting of *SS – CS* given as $y = (k \pm \Delta k)x + (k_0 \pm \Delta k_0)$

| plots | *Sample* | k | Δk | k₀ | Δk₀ | r | p |
|---|---|---|---|---|---|---|---|
| *SS – CS* | Whole | 0.96 | 0.24 | -6.22 | 0.42 | 0.62 | $3.14\times10^{-22}$ |
| *SS – CS* | Seyfert galaxies and FSRQs | 0.82 | 0.20 | -5.03 | 0.17 | 0.71 | $3.14\times10^{-17}$ |
| *SS – CS* | BL Lacs | 0.74 | 0.18 | -5.20 | 0.54 | 0.57 | $3.14\times10^{-20}$ |

To consider the form of relationship existing between the inverse Compton spectrum and both the synchrotron and Compton spectra, the scatter plots of *IC – SS* and *IC – CS* are shown in Fig. 5. For the *IC - SS* data in Fig. 5(a), there is a negative correlation for the individual subsamples, which upturns into a positive correlation on *IC –CS* plane. Three distinct groups of objects can be identified in each plot, namely, Seyfert galaxies, BL Lacs and FSRQs. These three groups of objects are slightly comparable on *IC - SS* and *IC – CS* planes with a kind of scaling factor on *SS* and *CS* though on a reverse direction. We interpret this to mean that similar processes give rise to the *IC - SS* anti-correlation and *IC – CS* positive correlation in the different groups at intrinsically different scales. The observed dichotomy between Seyfert galaxies and blazar samples is as a result of the differences in some intrinsic properties that vary in a sequence across the samples, which is consistent with the unification scheme. Tables 5a and 5b show the results of regression analyses of *IC –SS* and *IC –CS* data respectively. The slope *k*, intersection $k_0$, correlation coefficient *r* and chance probability *p* and their errors are all listed in the table. It is noteworthy that the correlations between the subsamples were individually and wholly considered in the analyses.

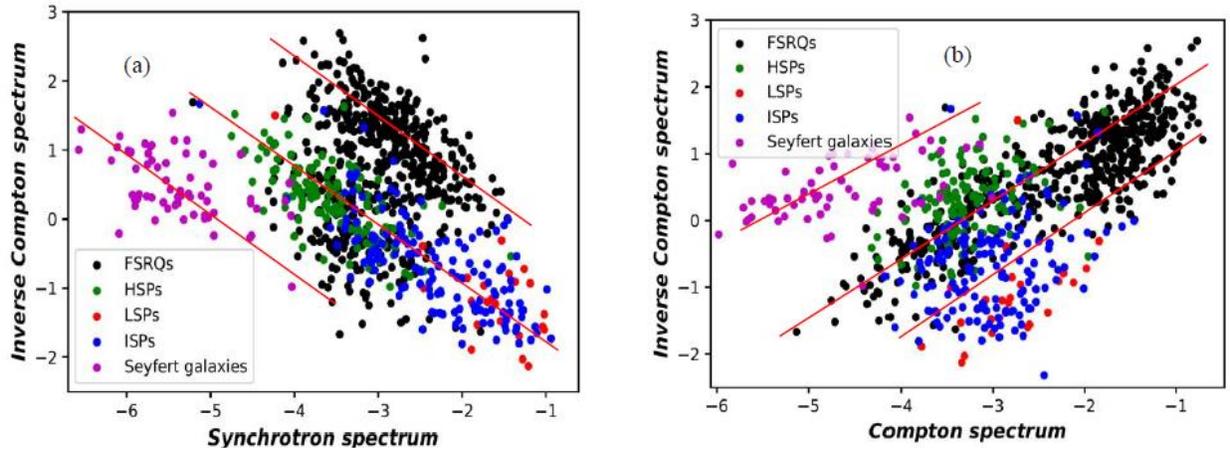

**Figure 5: Plot of inverse Compton spectrum against (a) synchrotron spectrum (b) Compton spectrum for FSRQs, Seyfert galaxies and BL Lac subclasses**

Table 5a: Results of linear regression fitting of *IC – SS* given as $y = (k \pm \Delta k)x + (k_0 \pm \Delta k_0)$

| plots | *Sample* | k | Δk | $k_0$ | $\Delta k_0$ | r | p |
|---|---|---|---|---|---|---|---|
| *IC – SS* | Whole | -0.93 | 0.31 | 2.17 | 0.08 | -0.65 | $3.22 \times 10^{-16}$ |
| *IC – SS* | Seyfert galaxies | -0.75 | 0.26 | 1.08 | 0.06 | -0.52 | $4.62 \times 10^{-13}$ |
| *IC – SS* | BL Lacs | -0.56 | 0.23 | 2.90 | 0.10 | -0.62 | $1.08 \times 10^{-17}$ |
| *IC – SS* | FSRQs | -0.56 | 0.32 | 2.90 | 0.10 | -0.68 | $7.05 \times 10^{-18}$ |

**Table 5a: Results of linear regression fitting of *IC –CS* given as** $y = (k \pm \Delta k)x + (k_0 \pm \Delta k_0)$

| plots | *Sample* | k | Δk | k₀ | Δk₀ | r | p |
|---|---|---|---|---|---|---|---|
| IC – CS | Whole | 1.14 | 0.34 | 0.72 | 0.38 | 0.57 | 3.02×10⁻¹⁷ |
| IC – CS | Seyfert galaxies | 2.03 | 0.20 | 0.22 | 0.26 | 0.56 | 1.00×10⁻¹⁴ |
| IC – CS | BL Lac subclasses | 1.25 | 0.23 | -4.32 | 0.30 | 0.58 | 4.05×10⁻¹⁶ |
| IC – CS | FSRQs | 1.30 | 0.16 | -5.01 | 0.20 | 0.61 | 2.04×10⁻¹⁵ |

## 4.3 Effects of Redshift on continuous spectra of our sample

The *SS – CS, IC – SS* and *IC – CS* significant correlations observed in section 4.2 may be due to redshift effect. In order to investigate for the real correlations, one has to eliminate the effects of luminosity distance on the sample by tracing out the common dependence of *SS, IC* and *CS* on redshift. This is done by performing the Spearmann's partial correlation analyses (e.g. Padovani et al. 1992; Odo & Ubachukwu, 2013) given as

$$r_{12,3} = \frac{r_{12} - r_{13}r_{23}}{\sqrt{(1-r_{13}^2)(1-r_{23}^2)}} \qquad (6)$$

where $r_{12}$ represents the correlation coefficient between $x_i$ and $x_j$, and $r_{ij,k}$ is the partial correlation coefficient between *xi* and *xj*, with *z* dependence removed; (*i, j, k* =1,2,3 = SS, IC, CS respectively). The Spearmann's test results are all shown in Tables 6a, 6b and 6c for *SS – CS, IC – SS* and *IC – CS* respectively. In these tables, the column (1) gives the plot, column (2) gives the class of the sample, column (3) is the correlation coefficient, column (4) is the correlation coefficient between two variables $x_i$ and $x_j$ with the redshift effect, column (5) is correlation coefficient between the variables $x_i$ and the redshift *z*, column (6) is the correlation coefficient with the redshift effect removed, column (7) gives the chance probability for the two distributions to come from the same population. Thus, we argue that the correlations are not driven by redshift effects, rather they are intrinsically induced.

**Table 6a:** *Regression analysis of SS – CS independent of redshift effect*

| plots | Sample | $r$ | $r_{ij}$ | $r_{jk}$ | $r_{ijk}$ | $p$ |
|---|---|---|---|---|---|---|
| $SS-CS$ | whole | 0.62 | 0.56 | 0.51 | 0.64 | $1.13\times10^{-15}$ |
| $SS-CS$ | Seyfer galaxies - FSRQs | 0.71 | 0.64 | 0.67 | 0.69 | $1.13\times10^{-14}$ |
| $SS-CS$ | BL Lacs | 0.57 | 0.53 | 0.50 | 0.58 | $1.13\times10^{-12}$ |

**Table 6b:** *Regression analysis of IC – SS independent of redshift effect*

| plots | Sample | $r$ | $r_{ij}$ | $r_{jk}$ | $r_{ijk}$ | $p$ |
|---|---|---|---|---|---|---|
| $IC-SS$ | whole | -0.63 | -0.67 | 0.63 | -0.62 | $3.10\times10^{-19}$ |
| $IC-SS$ | Seyfert galaxies - | -0.52 | -0.49 | -0.44 | -0.54 | $11.2\times10^{-18}$ |
| $IC-CS$ | BL Lacs | -0.62 | -0.57 | -0.61 | -0.63 | $4.43\times10^{-17}$ |
| $IC-SS$ | FSRQs | -0.68 | -0.59 | -0.64 | -0.65 | $1.65\times10^{-16}$ |

**Table 6c:** *Regression analysis of IC – CS independent of redshift effect*

| plots | Sample | $r$ | $r_{ij}$ | $r_{jk}$ | $r_{ijk}$ | $p$ |
|---|---|---|---|---|---|---|
| $IC-CS$ | whole | 0.57 | 0.53 | 0.60 | 0.55 | $2.30\times10^{-17}$ |
| $IC-CS$ | Seyfert galaxies - | 0.56 | 0.59 | 0.48 | 0.60 | $3.02\times10^{-19}$ |
| $IC-CS$ | BL Lacs | 0.58 | 0.62 | 0.52 | 0.56 | $2.33\times10^{-18}$ |
| $IC-CS$ | FSRQs | 0.61 | 0.66 | 0.63 | 0.63 | $8.09\times10^{-19}$ |

## 5 Discussion

Ever since Seyfert galaxies were discovered as jetted AGNs, there have been some debates as to whether these objects are fundamentally discrete subclass of AGNs or merely ow luminous objects under evolution (e.g. Osterbrock & Pogge, 1985). Several authors agree with the last assumption (see, e.g. Grupe 2000; Botte et al. 2004). The unification of relativistic jets provided further support for this supposition (see, Foschini 2011, 2012; 2014). In this paper, we collected a large sample of blazars and Seyfert galaxies, did some statistical work on them and compared the emission properties of these sources in the context of the unification scheme. On the basis of what we have found in this survey, with more sources and data, we found that, although Seyfert galaxies display some irregular observational differences with respect to the other radio-loud AGNs, the emission characteristics inferred from the current analysis (synchrotron, inverse Compton and Compton spectra) support the hypothesis that these sources are under development. It is a known assumption that the broadband SEDs of blazars have double bumps that usually peak at the radio to X-ray range and dominated by synchrotron emission (Marscher, 1998) while the second one is within the energy range of GeV - TeV and completely dominated by inverse Compton spectrum (Sambruna et. al. 1996). The high energy emission of blazars is largely modeled in terms of leptonic Compton up-scattering of soft synchrotron photons in the blazar jets, otherwise known as synchrotron-self Compton model (Marscher, 1992; 1998), or from external field photons, commonly referred to as External Compton (EC) model (Finke, 2013). Similarities in the distributions of parameters of jetted Seyfert galaxies and blazars observed in this study suggest that similar mechanisms take place in these sources. Our current results have remarkably shown from the comparison of the distributions of synchrotron, inverse Compton and Compton continuous spectra that FSRQs could be extreme version of Seyfert galaxy populations, thus, this implies to be in agreement with previous hypothesis that blazar unification can be extended to the low luminosity side of radio galaxy populations, believed to harbor powerful relativistic jets with extended radio structures. These radio galaxies are comparable to the jetted Seyfert galaxies, thus can be unified by invoking unification scheme through orientation (Pei et al. 2019, Iyida et al. 2020). Also, FSRQs have been reported to have the tendency of evolving into BL Lacs, becoming weak-lined objects by virtue of their increased continuum (Vagnetti et al., 1991) suggesting that Seyfert galaxies can be unified with the traditional radio-loud AGNs sample.

The unification scheme proposes that the FSRQs and BL Lacs are different expressions of the same physical process that vary only by bolometric luminosity (Ghisellini et al. 1998: Fossati et al. 1998). Consequently, there should be continuity in distributions of the continuous spectra of these objects. From the distributions of the continuous spectra of our sample, it is shown that these distributions can be extended to the jetted, radio-quiet Seyfert galaxies. These Seyfert galaxies are somewhat occupying the lowest regime while FSRQs occupy the highest regime of the distribution and BL Lacs being intermediate in the configuration. This is actually in agreement with evolutionary connection in which the AGNs may start off as Seyfert galaxy and grow in different emission spectra through BL Lacs into FSRQs suggesting that Seyfert galaxies are young, jetted AGNs that are still under evolution (Singh & Chand, 2018). On the other hand, the distribution of Compton spectrum does not apparently show that Seyfert galaxies are observed at significantly low energy component from other subclasses of the AGNs. Thus, it can be argued from the distributions of Compton spectrum that orientation effect may not be the major difference between Seyfert galaxies and blazar subclasses. In addition, if the population arguments are correct, then one can expect that the distributions of the synchrotron spectrum for blazars and galaxies should be from the same parent population. The *K-S* test of the distributions of our sample shows that these jetted extragalactic sources could be unified.

Also, within the context of the unification scheme, the apparent order observed between the Seyfert galaxies, BL Lacs and FSRQs are attributable to a unification scheme for all jetted AGNs. Our results indicate that there are inherent differences in the location of our samples in *SS - CS* plane. The low Compton spectrum of Seyfert galaxies has an important implication. It is observed that this Compton spectrum is low for Seyfert galaxies but comparable to FSRQs than BL Lacs. The seyfert galaxies are evidently less luminous than BL Lacs. This could be because the later can emit more at X-ray/γ-ray frequencies than the former since the Compton spectrum peaks in the X-rays/ γ-ray region (see, Fig. 5b), and thus, BL Lacs are likely to be found in X-ray surveys (Padovani & Giommi 1995). Figure 5a shows that Seyfert galaxies are at the low synchrotron spectrum which agrees with previous works that these sources are dominant at the low energy components (see, e.g. Foschini et al., 2016). While at the radio to UV/X-ray frequencies, Seyfert galaxies and blazar subclasses diverge from each other at this spectrum with BL Lac objects moving to larger synchrotron spectrum, indicating a different origin of the emission (synchrotron for BL Lacs) and inverse Compton spectrum for Seyfert galaxies. The variations in these continuous spectra between FSRQs and BL Lacs can be understood in terms of their emission mechanism of the unified scheme such that a common factor that change linearly between the subclasses is responsible for the variation. The interpretation is that the main mechanism that is generating the broadband emissions in our sample are the same, though, they differ systematically among the AGNs subclasses (see, e.g. Chen et al. 2016). Thus, this indicates an evolutionary link among these extragalactic sources which is in agreement with blazar unification scheme as earlier suggested by Fossati et al. (1998) and Ghisellini et al. (1998).

We observed the existence of a selection of high frequency-peaking FSRQs (about 14 %) that have the same value of parameter in $SS - CS$ and $IC - SS$ planes. These few FSRQs are occupying the parameter space in a way that agrees with the unification scheme for radio galaxies, BL Lacs and FSRQs, (see e.g. Iyida et al. 2020; Odo & Aroh, 2020). Actually, the sequence from Seyfert galaxy at low $SS - CS$ plane to FSRQs at high $SS - CS$ plane is in agreement with the evolution based unification scheme for Seyfert galaxy - BL Lacs - FSRQs unification such that at low synchrotron and inverse Compton energies, jetted AGNs are Seyfert galaxies though can metamorphose through BL Lac subclasses to FSRQs. So, the existence of the few FSRQs that share similar parameter space with the BL Lacs in the current sample does not change the evolutionary development from Seyfert galaxy to FSRQs through the BL Lac subclasses of blazars, but supports the unification scheme for Seyfert galaxies -BL Lac - FSRQs through evolution. Probably, this handful of FSRQs that are masquerading as BL Lacs are noted since their existence are gradually attracting attention in recent investigations (e.g. Giommi et al. 2013; Padovani et al. 2019; Odo & Aroh, 2020). Therefore, it is argued that these class of FSRQs could possibly be a special subgroup of high synchrotron-peaking BL Lacs.

## 6 Conclusion

We have investigated the relationship between radio-quiet Seyfert galaxies and subclasses of jetted radio-loud AGNs using observed properties of a sample of extragalactic sources. We show from the distributions of the computed continuous spectra that Seyfert galaxies form the tail of distributions of the jetted AGNs which is consistent with the scenario that Seyfert galaxies are young growing jetted objects. Furthermore, distributions of the objects on $SS - CS, IC - SS$ and $IC - CS$ planes show that Seyfert galaxies, BL Lacs and FSRQs possess similar evolutionary histories in similar environments. There is a significant positive correlation ($r \sim 0.60$) on $SS - CS$ and *IC – CS* data in each individual subsample and this upturns into anti-correlation (*r* ~ -0.60) on *IC – SS* plane. These significant correlations imply that these extragalactic sources can be unified based on their broadband emission properties.

**Acknowledgement**

We thank the anonymous referees for valuable comments and suggestions which helped to improve the manuscript.